\documentclass{arxivpubl}

\usepackage[T1]{fontenc}
\usepackage{dfadobe}  
\usepackage{algorithm}
\usepackage{algorithmic}
\usepackage{xcolor}
\usepackage{pgf}
\usepackage[percent]{overpic}
\usepackage{cite}  

\BibtexOrBiblatex

\electronicVersion
\PrintedOrElectronic
\usepackage{graphicx}  

\title[Real-Time Path-Guiding Based on Parametric Mixture Models]%
      {Real-Time Path-Guiding Based on Parametric Mixture Models\vspace{-4ex}}

\author[Mikhail Derevyannykh]
{\parbox{\textwidth}{\centering Mikhail\, Derevyannykh\\ Moscow State University, Eagle Dynamics}}

\begin{document}
\teaser{
  \begin{minipage}[b]{0.48\textwidth}
  \begin{overpic}[width=\linewidth]{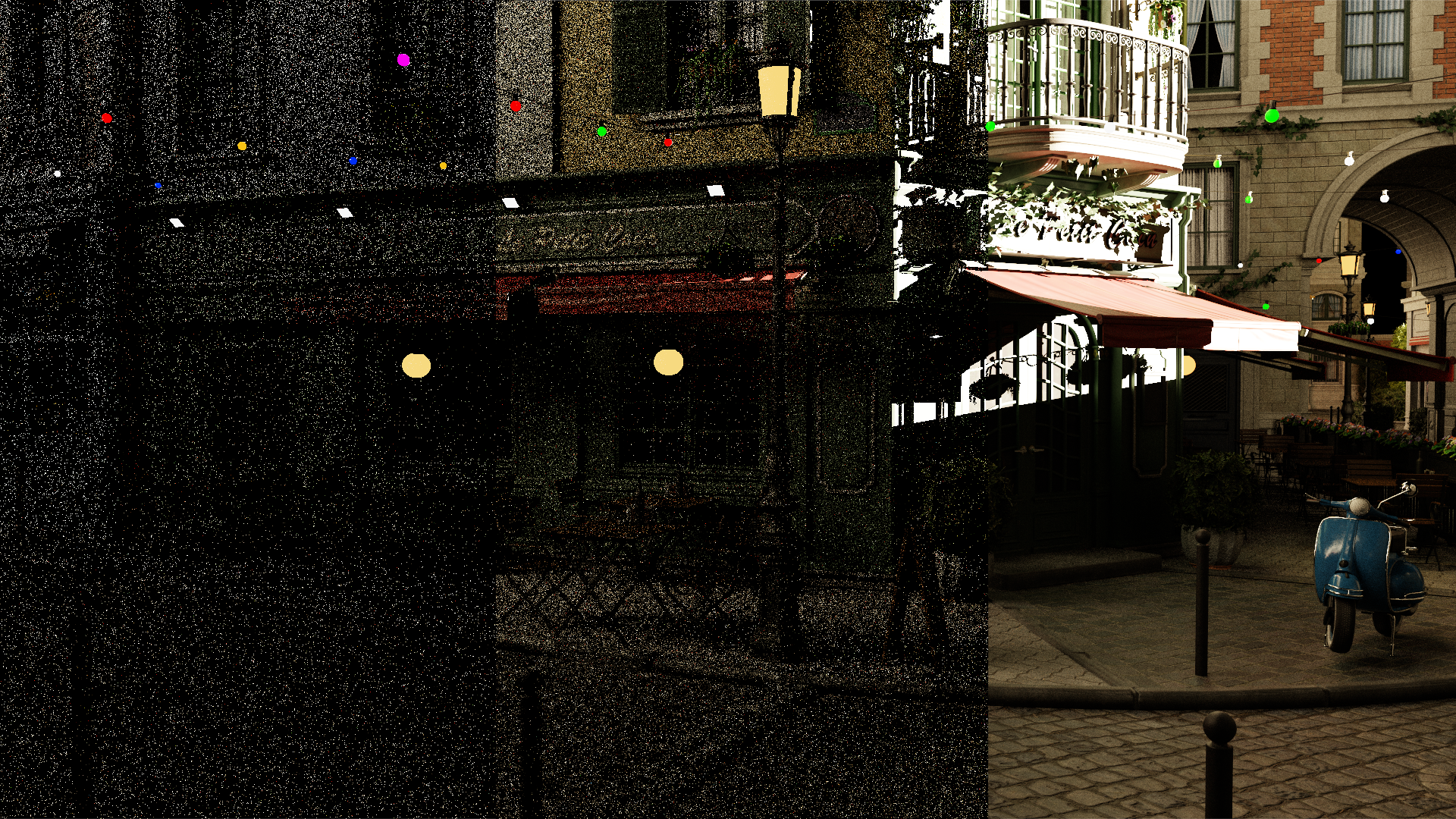}
  \put (34.5,2.5) {\textcolor{white}{\footnotesize PT + PG (Ours) 40.5 ms}}
 \put (3, 2.5) {\textcolor{white}{\footnotesize Path-Tracing 42.0 ms}}
 \put (73.5,2.5) {\textcolor{white}{\footnotesize MC Reference}}
  \end{overpic}
  \end{minipage}
  \begin{minipage}[b]{0.48\textwidth}
  \begin{overpic}[width=\linewidth]{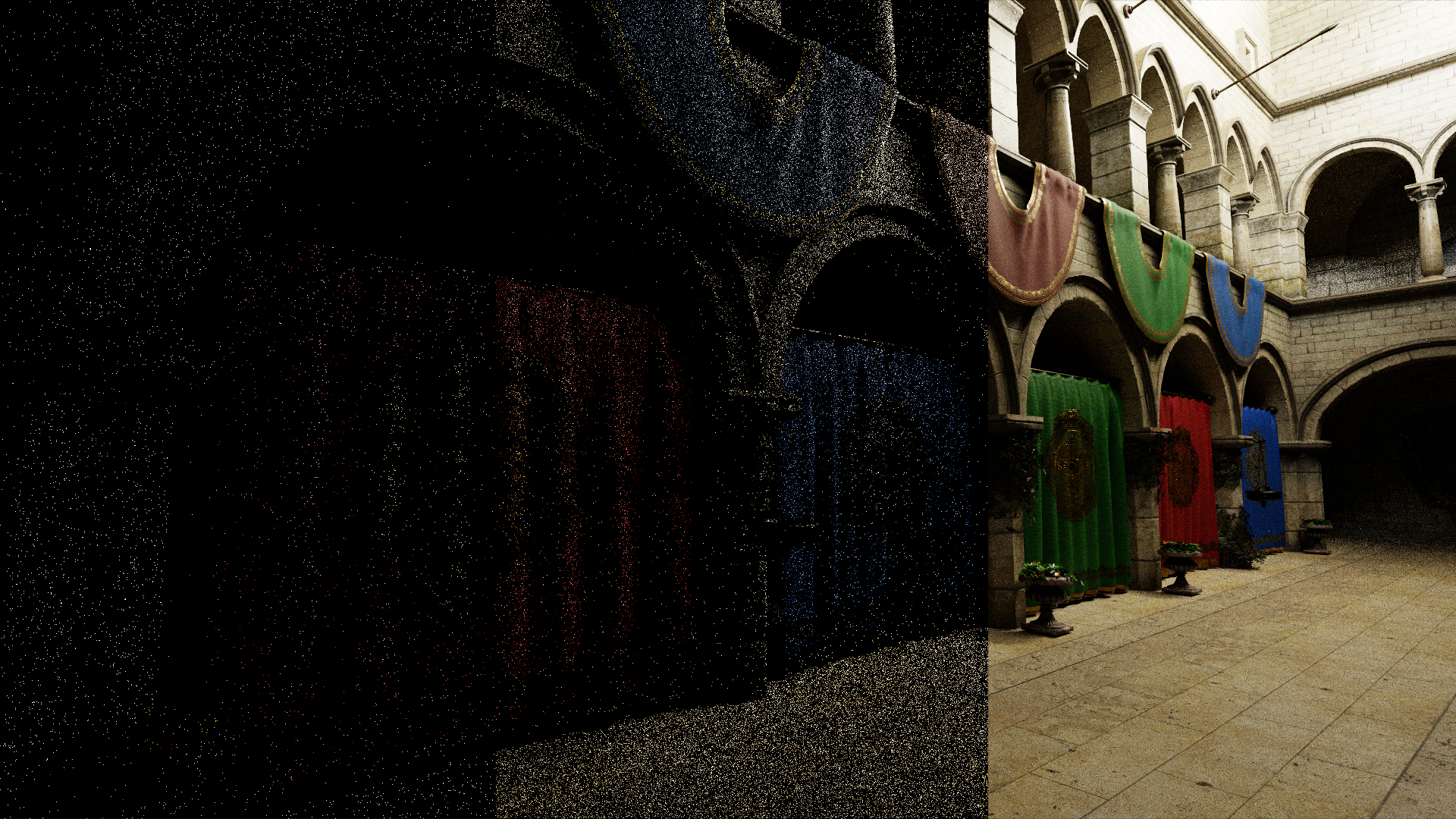}
  \put (34.5,2.5) {\textcolor{white}{\footnotesize PT + PG (Ours) 35.0 ms}}
 \put (3, 2.5) {\textcolor{white}{\footnotesize Path-Tracing 37.4 ms}}
 \put (73.5,2.5) {\textcolor{white}{\footnotesize MC Reference}}
 \end{overpic}
  \end{minipage}
  \centering
   \caption{
           Frames of the Bistro scene (left) and the Sponza scene (right) with complex light scenarios are rendered with 1 spp using Path-Tracing (PT) and our Path-Guiding (PG) solution, and also with 2048 spp for reference. Notice the decrease in overall noise level for diffuse and specular surfaces, the ability of our algorithm to capture high-frequency indirect shadows just relying on 1 spp. Also, there is a slight performance boost.
           The images were rendered at 1080p resolution on a high-end desktop machine with RTX 2070 using Falcor framework \cite{Falcor}. }
 \label{fig:teaser}
}
\maketitle
\begin{abstract}
       Path-Guiding algorithms for sampling scattering directions can drastically decrease the variance of Monte Carlo estimators of Light Transport Equation, but their usage was limited to offline rendering because of memory and computational limitations. We introduce a new robust screen-space technique that is based on online learning of parametric mixture models for guiding the real-time path-tracing algorithm. It requires storing of 8 parameters for every pixel, achieves a reduction of \textnormal{\reflectbox{F}LIP} metric up to 4 times with 1 spp rendering. Also, it consumes less than 1.5ms on RTX 2070 for 1080p and reduces path-tracing timings by generating more coherent rays by about 5\% on average. Moreover, it leads to significant bias reduction and a lower level of flickering of SVGF output.   
\end{abstract}  
\section{Introduction}
The main goal of path-tracing is to calculate solution of Light Transport Equation \cite{LTE}:
\[L_{o}(x, \omega_{o}) = L_{e}(x, \omega_{o}) + \int_{H^{+}} L_{i}(x, \omega_{i})f_{r}(x,\omega_{i},\omega_{o})cos(\theta_{i}) d\omega_{i}  (1) \]
Here \(L_{e}(x, \omega_{o})\) is the radiance emitted from a point \(x\) in a direction \(\omega_{o}\), \(L_{i}(x, \omega_{i})\) - is the incoming radiance from \(\omega_{i}\), \(f_{r}\) - BRDF and \(\theta_{i}\) - angle between the surface normal at \(x\) and \(\omega_{i}\). Using Monte Carlo simulation method for solving it \cite{LTEVeach} leads to high variance. Usually Importance Sampling techniques are used for sampling \(\omega_{i}\) direction to decrease the variance, but they only take into account BRDF term of Eq. (1) and lead to unpleasant results in complex light cases. That's why we need to guide the process of generating scattering direction based on \(L_{i}(x, \omega_{i})cos(\theta_{i})\) \cite{PMM} \cite{MullerPractical} term, too, especially for real-time ray-tracing with limited samples budget.

In this paper, we suggest to store parameters of parametric mixture models \cite{PMM} for every pixel in additional 2D textures \(\Gamma\) and use them for guiding. Mixture model \(D_{w, h}\) for a pixel \(p_{w, h}\) is based on BRDF distribution of its surface \(x_{w, h}\) and 2D Gaussian Distribution:
\[D_{w, h}(\omega_{i}) = \pi_{w,h} N(M^{-1}(\omega_{i}),\mu_{w,h},\Sigma_{w,h})+\pi_{w,h}^{-1}f_{r}(x_{w, h},\omega_{i},\omega_{o}) (2)\]
where \(\mu_{w,h}\) - mean, \(\Sigma_{w,h}\) - covariance matrix, mixing coefficient \(\pi_{w, h}\), \(\pi_{w,h}^{-1} =1-\pi_{w,h}\)  and \(M\) - Shirley and Chiu area preserving mapping \cite{MAPPING} from 2D unit square to 3D unit hemisphere, \(N\) - pdf of normal distribution. This technique ensures supporting arbitrary scenes, the ability to train models iteratively by re-projecting data from old frames, and robust Multiple Importance Sampling \cite{MIS} for achieving results no worse than using default BRDF sampling. Our solution leads to a decrease in the overall level of noise (Figure 1) and doesn't impose any technical limitations on scenes materials, geometry and etc.
\section{Previous Work}
\begin{figure}[htb]
  \centering
  \begin{overpic}[width=1.0\linewidth]{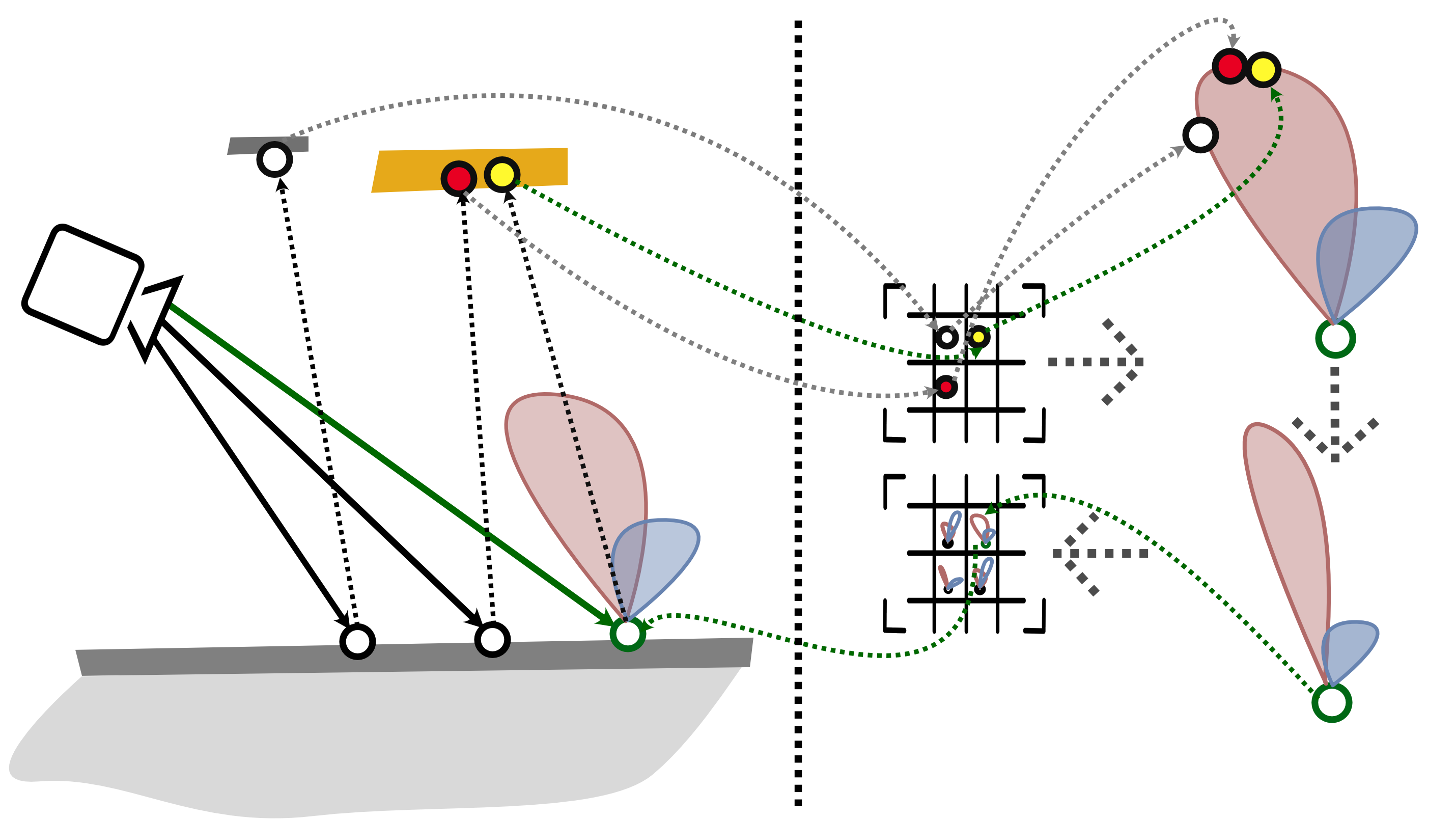}
  \put (16,53) {\textcolor{black}{\footnotesize Path Tracing Pass}}
  \put (27,47) {\textcolor{gray}{\scriptsize Bright surface}}
  \put (10,48) {\textcolor{gray}{\scriptsize Dark surface}}
  \put (4.0,43.0) {\textcolor{gray}{\rotatebox{-25}{\scriptsize Camera}}}
  \put (53,8.5) {\textcolor{gray}{\rotatebox{0}{\parbox{7em}{\scriptsize  \centerline{Frame Buffer \textcolor{black}{\(\Gamma\)}}}}}}
  \put (53,40.5) {\textcolor{gray}{\rotatebox{0}{\parbox{7em}{\scriptsize  \centerline{Frame}
  \centerline{Buffer}
  \centerline{\textcolor{black}{\(\Pi\)}}}}}}
  \put (95,15.0) {\textcolor{black}{\rotatebox{90}{\parbox{7em}{\scriptsize  \centerline{ Expectation}
  \centerline{Maximization}}}}}
  \put (7,64.0) {\textcolor{black}{\rotatebox{-45}{\scriptsize  \centerline{\(N(\mu_{1, 1},\Sigma_{1, 1})\)}}}}
  \put (13.5,55.0) {\textcolor{black}{\rotatebox{-45}{\scriptsize  \centerline{\(f_{r}(x_{1, 1})\)}}}}
  \put (-6,9.0) {\textcolor{black}{\scriptsize  \centerline{\(x_{1, 1}\)}}}
  \put (-35,35.2) {\textcolor{black}{\scriptsize  \centerline{\(p_{1, 1}\)}}}
  \put (-8.5,25.2) {\textcolor{black}{\scriptsize  \centerline{\(\omega_{i}\)}}}
  \put (-12.0,13.2) {\textcolor{black}{\scriptsize  \centerline{\(-\omega_{o}\)}}}
  \put (-16,9.0) {\textcolor{black}{\scriptsize  \centerline{\(x_{0, 0}\)}}}
  \put (-25,9.0) {\textcolor{black}{\scriptsize  \centerline{\(x_{0, 1}\)}}}
  \put (-12,44.3) {\textcolor{black}{\scriptsize  \centerline{\(y_{1, 1}\)}}}
\put (-22,44.3) {\textcolor{black}{\scriptsize  \centerline{\(y_{0, 0}\)}}}
\put (-28,44.3) {\textcolor{black}{\scriptsize  \centerline{\(y_{0, 1}\)}}}
  \put (69,1) {\textcolor{black}{\footnotesize Training Pass}}
  \end{overpic}
  %
  \caption{\label{fig:firstExample}
           Firstly, we trace a ray from pixel \(p_{1, 1}\) to scene with direction \(-\omega_{o}\) and get an intersection in \(x_{1, 1}\) point respectively. Then algorithm loads parameters of mixture models from the framebuffer \(\Gamma(1, 1)\) for sampling first scattering direction \(\omega_{i}\), traces a ray from \(x_{1, 1}\) to \(\omega_{i}\) and intersects some surface in \(y_{1,1}\) point. We save \(y_{1,1}\) and estimated \(L(x_{1,1}, \omega_{i})\) as Virtual Point Light in \(\Pi(1,1)\) framebuffer. In the second training stage we estimate luminance of \(L_{i}(x_{1, 1}, \omega_{i})f_{r}(x_{1, 1},\omega_{i},\omega_{o})cos(\theta_{i})\) by accessing \(\Pi\), use it and \(\omega_{i}\) for training mixture model of \(x_{1,1}\) by Expectation-Maximization (EM) algorithm in on-line fashion and save updated params in \(\Gamma(1, 1)\). Also we use VPL of close pixels (\(p_{0, 0}\), \(p_{0, 1}\) in this example) for faster training.}
           \vspace{-3ex}
\end{figure}
The majority of previous works about path guiding were based on partitioning rendering scenes by some complex spatial structure on small cells and iteratively estimating spherical or hemispherical incoming radiance distribution for their internal surfaces by several approaches \cite{MullerPractical}\cite{PMM}\cite{VarianceAware}. Authors of \cite{PMM} rely on parametric mixture models for representing the incoming radiance, and they developed an algorithm for estimating model parameters in an online fashion with the support of weighted samples by radiance values. We use the same scheme with adaptive BRDF sampling and efficient implementation for real-time rendering based on spatial sampling Virtual Point Lights (VPL) \cite{VPL} as in Restir GI algorithm \cite{RestirGI}.
\section{Path-Guiding Pipeline}
We construct our solution based on the default path tracing algorithm with a subsequent training pass. It doesn't need any additional precomputations and complex volumetric data structures. Figure 2 illustrates the overall concept of our algorithm.
\subsection{Path tracing pass}
Algorithm inference default path tracing using rasterized G-Buffer, but with generating first scattering direction \(\omega_{i}\) based on the learned distribution of point \(x_{0}\). Parameters for constructing this distribution are requested from texture \(\Gamma\) with NN filtering, based on current pixel coordinates \(w, h\) and motion vectors. For estimating mean and co-variance matrix in online fashion we only store and update \(E(X), E(Y), E(X^{2}), E(Y^{2}), E(XY)\) weighted by radiance's luminance moments of the sampled distribution, where \([X, Y] =  M^{-1}(\omega_{i})\) and \(E\) - expected value. In cases of invalid history (new visible geometry without trained distribution) we initialize new parametric models with \(\pi_{w,h}\) = 0.05, \(k\) = 0 (it is a number of training epochs) and sample direction from BRDF.

For generating scattering direction, in case of sampling uniform random variable \(\zeta\) < \(\pi_{w, h}\) we employ simulation of two \(N(0, 1)\) random variables by Box-Muller method and transforming them to \(N(\mu_{w, h},\Sigma_{w, h})\) distribution with \(M\) mapping, in other - use BRDF.
In the end of pass the algorithm also saves estimated \(L_{i}(x_{w, h},\omega_{i})\) and position of next surface interaction \(y_{w, h}\) in additional texture \(\Pi\) for representing VPL.

For achieving stable re-projecting mixture models parameters from an old frame to a new one we use some heuristical stopping weighting based on depth and normal differences, like in \cite{SVGF} work. Also, it's important to take into account the difference between the hemispherical space of re-projected distribution and \(x_{w, h}\) surface space.
\subsection{Training pass}
Expectation-Maximization (EM) algorithm iteratively trains parametric models to estimate distribution \(D_{w, h}(\omega_{i}) \approx (L_{i}(x, \omega_{i})f_{r}(x,\omega_{i},\omega_{o})cos(\theta_{i})\) for every pixel \(p_{w, h}\) based on incoming radiance \(L_{i}(x_{w, h},\omega_{i})\) from \(y_{w, h}\) and current fixed \(\omega_{o}\). For running E-Step it's supposed to fetch material parameters from G-Buffer and estimate posterior probability \(p\) of sampling \(\omega_{i}\) from BRDF and from \(N([x, y], \mu_{w, h},\Sigma_{w, h})\) distribution where \([x, y] = M^{-1}(\omega_{i})\). In M-step algorithm calculates new estimation of moments and other parameters of model relying on some sort of temporal "filtering" using number of training epoch \(k\). 

Also, we suggest using VPLs from neighboring pixels for accelerating convergence, because it does not lead to any bias. For training distribution we only use samples generated by BRDF strategy, because relying on all samples leads to \(\pi_{w,h} \approx 1.0\) and biased results in our experiments. For the same reasons we clamp \(\pi_{w, h}\) value to be in range [0.05,0.95].
\begin{figure*}[htbp]
\begin{minipage}[b]{0.245\textwidth}
  \begin{overpic}[width=\linewidth]{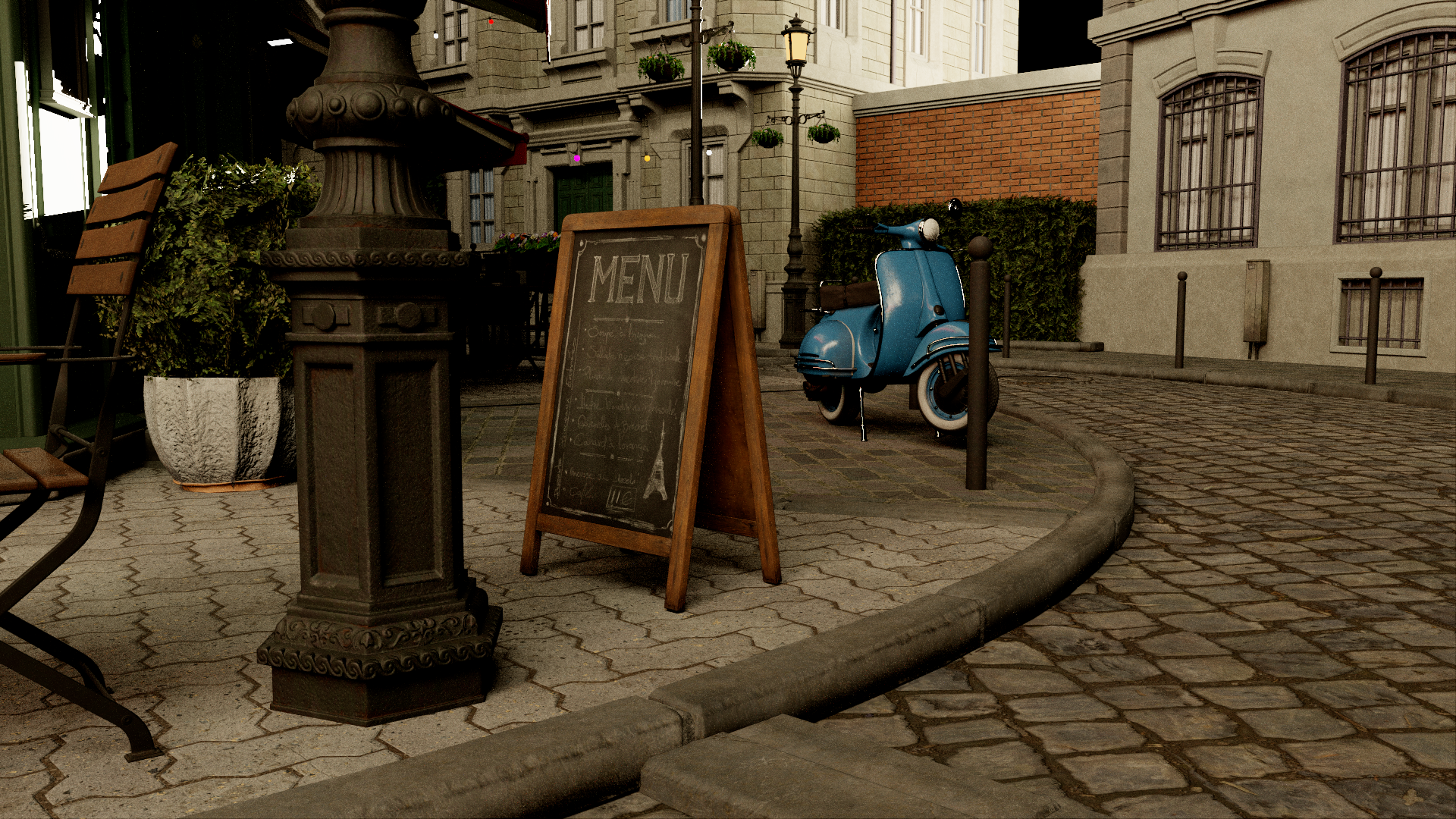}
  \put (-7, -25) {\rotatebox{90}{  \centerline{\textcolor{black}{\footnotesize Bistro}}}}
 \put (4, 2.5) {\textcolor{white}{\footnotesize Reference}}
  \end{overpic}
  \end{minipage}
  \begin{minipage}[b]{0.245\textwidth}
 \begin{overpic}[width=\linewidth]{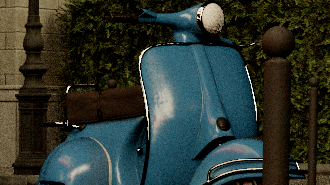}
 \put (70,2.5) {\textcolor{white}{\footnotesize \reflectbox{F}LIP: 0.0}}
 \put (4, 2.5) {\textcolor{white}{\footnotesize Reference}}
 \end{overpic}
  \end{minipage}
  \begin{minipage}[b]{0.245\textwidth}
  \begin{overpic}[width=\linewidth]{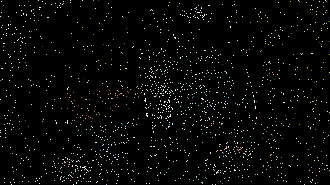}
  \put (70,2.5) {\textcolor{white}{\footnotesize \reflectbox{F}LIP: 0.47}}
 \put (4, 2.5) {\textcolor{white}{\footnotesize PT}}
 \end{overpic}
 \end{minipage}
 \begin{minipage}[b]{0.245\textwidth}
 \begin{overpic}[width=\linewidth]{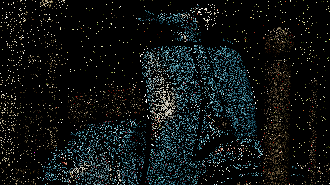}
  \put (70,2.5) {\textcolor{white}{\footnotesize \reflectbox{F}LIP: 0.37}}
 \put (4, 2.5) {\textcolor{white}{\footnotesize PG (Ours)}}
 \end{overpic}
  \end{minipage}
  \\[0.25cm]
  \begin{minipage}[b]{0.245\textwidth}
  \begin{overpic}[width=\linewidth]{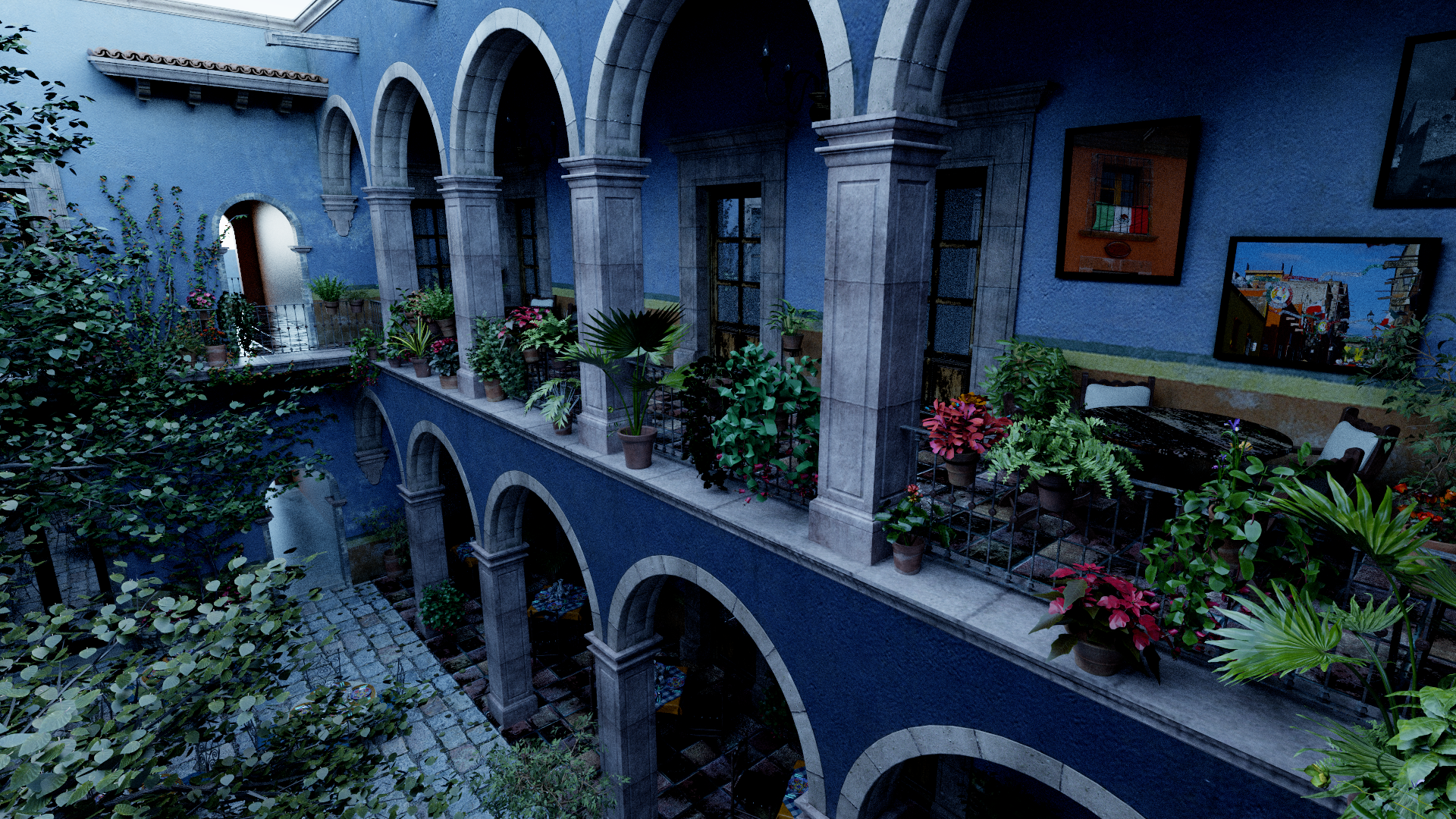}
  \put (-7, -25) {\rotatebox{90}{  \centerline{\textcolor{black}{\footnotesize San Miguel}}}}
 \put (4, 2.5) {\textcolor{white}{\footnotesize Reference}}
  \end{overpic}
  \end{minipage}
  \begin{minipage}[b]{0.245\textwidth}
 \begin{overpic}[width=\linewidth]{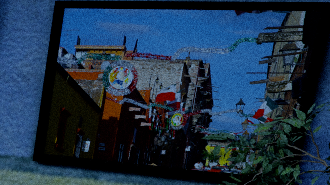}
 \put (70,2.5) {\textcolor{white}{\footnotesize \reflectbox{F}LIP: 0.0}}
 \put (4, 2.5) {\textcolor{white}{\footnotesize Reference}}
 \end{overpic}
  \end{minipage}
  \begin{minipage}[b]{0.245\textwidth}
  \begin{overpic}[width=\linewidth]{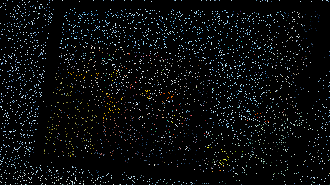}
  \put (70,2.5) {\textcolor{white}{\footnotesize \reflectbox{F}LIP: 0.43}}
 \put (4, 2.5) {\textcolor{white}{\footnotesize PT}}
 \end{overpic}
 \end{minipage}
 \begin{minipage}[b]{0.245\textwidth}
 \begin{overpic}[width=\linewidth]{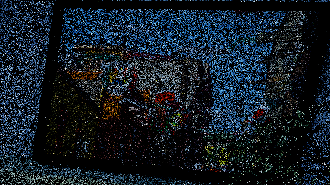}
  \put (70,2.5) {\textcolor{white}{\footnotesize \reflectbox{F}LIP: 0.29}}
 \put (4, 2.5) {\textcolor{white}{\footnotesize PG (Ours)}}
 \end{overpic}
  \end{minipage}
  \\[0.25cm]
  \begin{minipage}[b]{0.245\textwidth}
  \begin{overpic}[width=\linewidth]{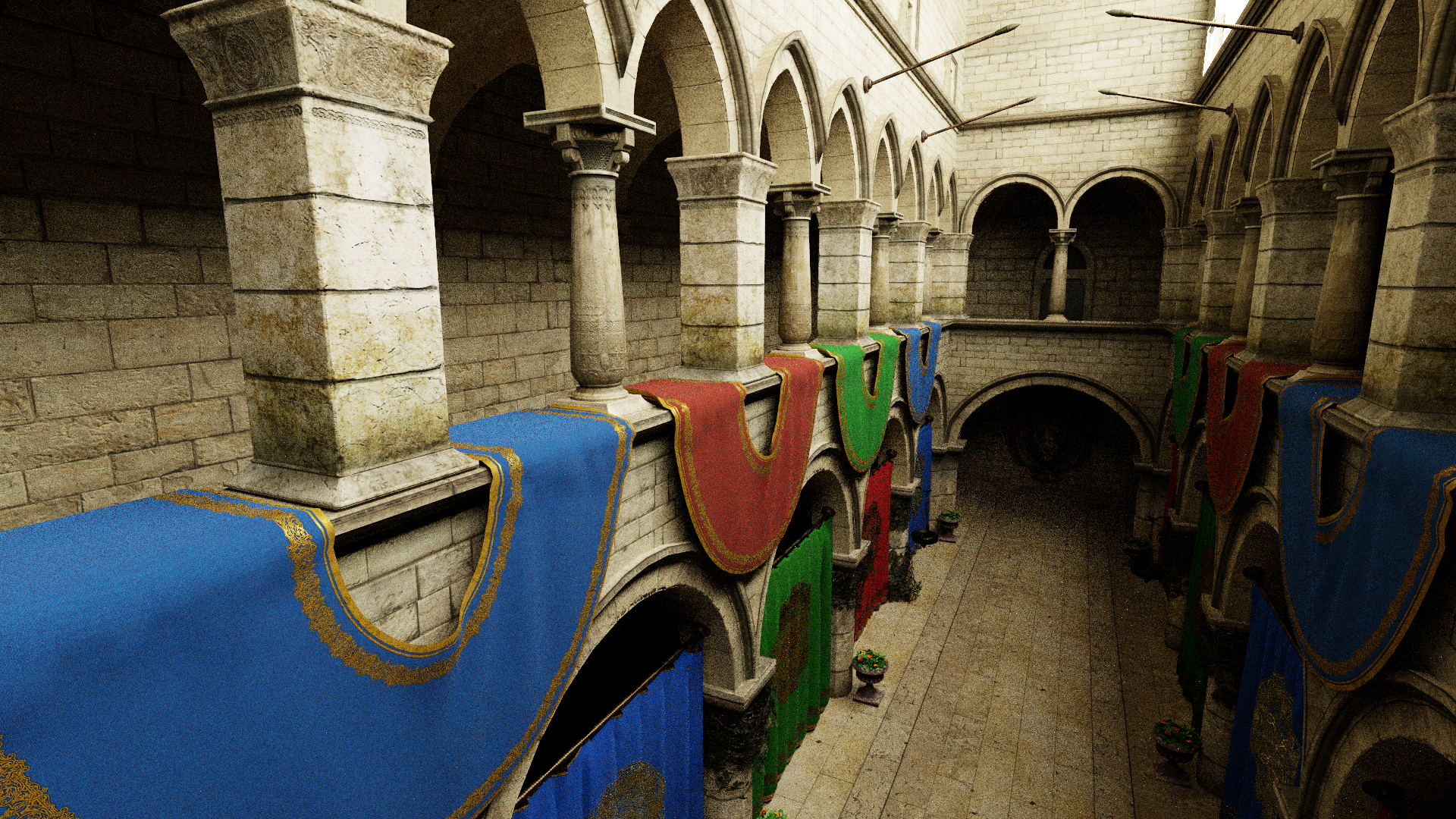}
  \put (-7, -25) {\rotatebox{90}{  \centerline{\textcolor{black}{\footnotesize Sponza}}}}
 \put (4, 2.5) {\textcolor{white}{\footnotesize Reference}}
  \end{overpic}
  \end{minipage}
  \begin{minipage}[b]{0.245\textwidth}
 \begin{overpic}[width=\linewidth]{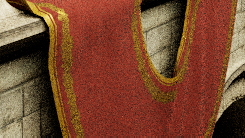}
 \put (70,2.5) {\textcolor{white}{\footnotesize \reflectbox{F}LIP: 0.0}}
 \put (4, 2.5) {\textcolor{white}{\footnotesize Reference}}
 \end{overpic}
  \end{minipage}
  \begin{minipage}[b]{0.245\textwidth}
  \begin{overpic}[width=\linewidth]{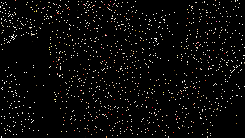}
  \put (70,2.5) {\textcolor{white}{\footnotesize \reflectbox{F}LIP: 0.71}}
 \put (4, 2.5) {\textcolor{white}{\footnotesize PT}}
 \end{overpic}
 \end{minipage}
 \begin{minipage}[b]{0.245\textwidth}
 \begin{overpic}[width=\linewidth]{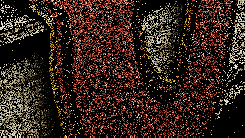}
  \put (70,2.5) {\textcolor{white}{\footnotesize \reflectbox{F}LIP: 0.47}}
 \put (4, 2.5) {\textcolor{white}{\footnotesize PG (Ours)}}
 \end{overpic}
  \end{minipage}
  \\[0.25cm]
  \begin{minipage}[b]{0.245\textwidth}
  \begin{overpic}[width=\linewidth]{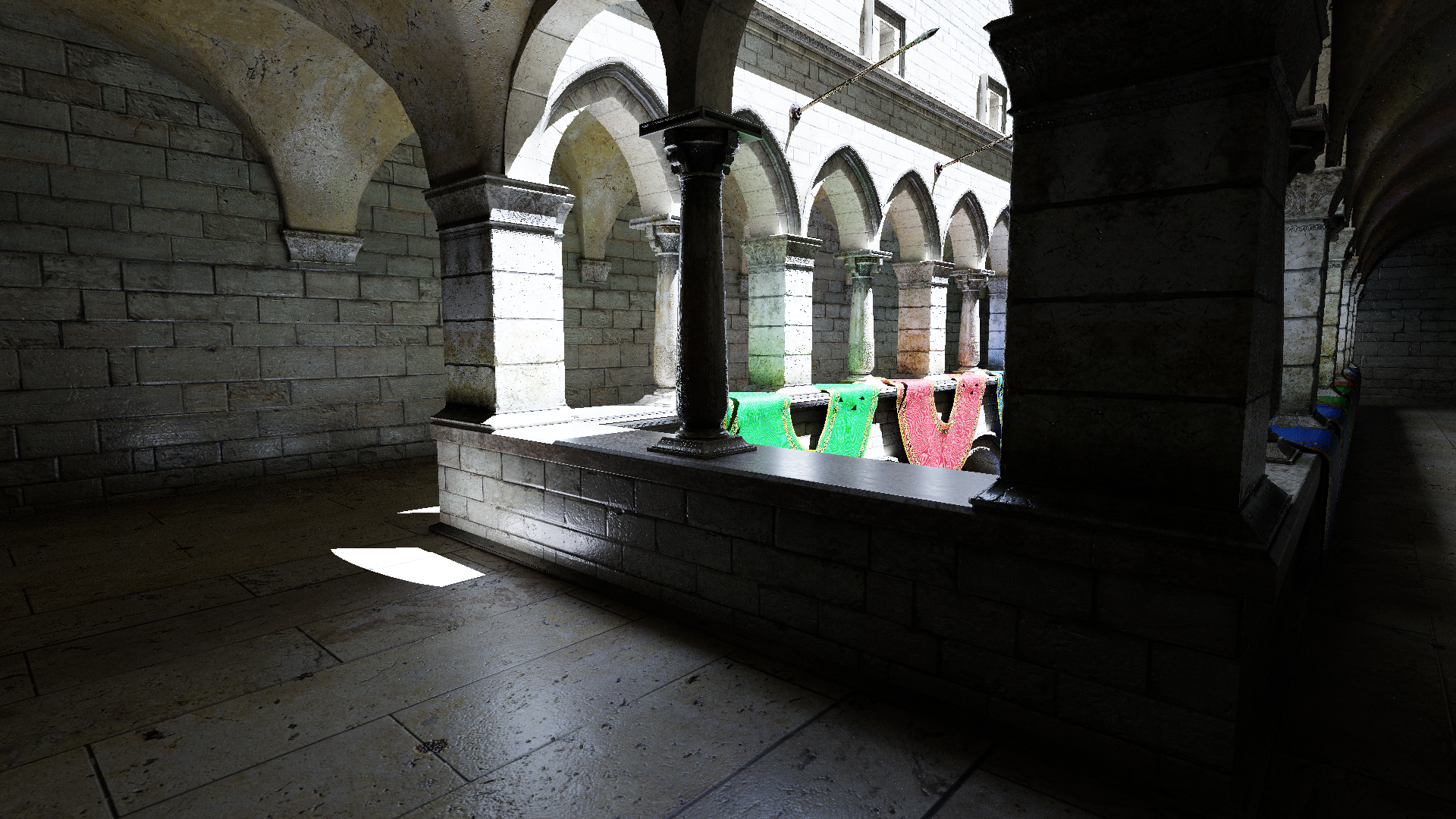}
  \put (-7, -25) {\rotatebox{90}{  \centerline{\textcolor{black}{\footnotesize Glossy Sponza}}}}
 \put (4, 2.5) {\textcolor{white}{\footnotesize Reference}}
  \end{overpic}
  \end{minipage}
  \begin{minipage}[b]{0.245\textwidth}
 \begin{overpic}[width=\linewidth]{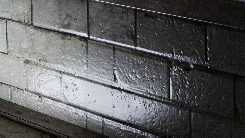}
 \put (70,2.5) {\textcolor{white}{\footnotesize \reflectbox{F}LIP: 0.0}}
 \put (4, 2.5) {\textcolor{white}{\footnotesize Reference}}
 \end{overpic}
  \end{minipage}
  \begin{minipage}[b]{0.245\textwidth}
  \begin{overpic}[width=\linewidth]{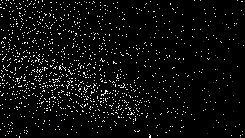}
  \put (70,2.5) {\textcolor{white}{\footnotesize \reflectbox{F}LIP: 0.57}}
 \put (4, 2.5) {\textcolor{white}{\footnotesize PT}}
 \end{overpic}
 \end{minipage}
 \begin{minipage}[b]{0.245\textwidth}
 \begin{overpic}[width=\linewidth]{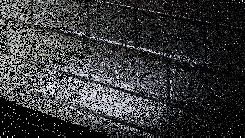}
  \put (70,2.5) {\textcolor{white}{\footnotesize \reflectbox{F}LIP: 0.31}}
 \put (4, 2.5) {\textcolor{white}{\footnotesize PG (Ours)}}
 \end{overpic}
  \end{minipage}
  \caption{\label{fig:firstExample}
           These comparisons demonstrate the effectiveness of our path-guiding algorithm for different scenes with zoomed render patches relying on 1 spp with trained mixtures. It decreases the variance of LTE (1) estimator for specular materials with normal maps (Bistro and Glossy Sponza), scenes with low-frequency environmental lighting (San Miguel), diffuse materials with high-frequency local indirect shadows, and complex sampling directional lighting (Sponza). Also we show \textnormal{\reflectbox{F}LIP}\cite{FLIP} metric error for numerical comparison.}
 \end{figure*}
\section{Results and Discussion}
We evaluate our new path-guiding algorithm on a set of different scenes: Sponza, Bistro Exterior, SanMiguel, Glossy Sponza based on the Falcor rendering framework \cite{Falcor}.
\subsection{Comparison with Path-Tracing}
We compare our solution only to default path-tracing because all other Path-Guiding techniques don't support real-time GPU-based rendering and require additional precomputations.
As can be seen in Figure 3 our algorithm drastically decreases the level of noise for all testing scenes. It allows us to distinguish indirect local shadows, high-frequency specular reflections which are caused by employing normal maps, only by relying on 1 sample per pixel. Also, we want to notice that our technique decrease variance of the MC estimator for low-frequency environmental lighting - San Miguel scene, too. But, unfortunately, path guiding can't decrease the noise level in rendering complex meshes like bushes in the Bistro frame behind the scooter, because there are a lot of self-shadowing effects and that's why it is hard to represent effectively incoming lighting distribution just relying on one 2D Normal Distribution and BRDF.

Real-time rendering community usually apply SVGF denoiser \cite{SVGF} for achieving stable rendering results. We suggest employing the path-guiding technique, because, as can be seen in Figure 4, it drastically improves the robustness of the SVGF algorithm. Also, our technique decreases the level of flickering and recover lost indirect shadows and lighting. For a demonstration of this combination of techniques please watch the videos from supplementary materials. But we rendered several frames of the static scene with disabled jittering of pixel centers using PT+SVGF and PG+SVGF, constructed plots with estimated temporal error (MSE between every consequent frame) for proving our statements about flickering level - see Figure 4.
\begin{figure}[h]
\begin{minipage}[b]{0.155\textwidth}
  \begin{overpic}[width=\linewidth]{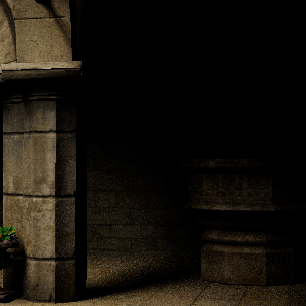}
 \put (4, 2.5) {\textcolor{white}{\footnotesize Reference}}
  \end{overpic}
  \end{minipage}
  \begin{minipage}[b]{0.155\textwidth}
 \begin{overpic}[width=\linewidth]{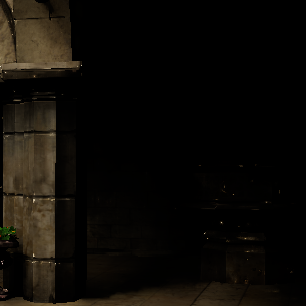}
 \put (53,2.5) {\textcolor{white}{\footnotesize \reflectbox{F}LIP: 0.13}}
  \put (51.5, 89) {\textcolor{white}{\footnotesize PT +SVGF}}
 \end{overpic}
  \end{minipage}
  \begin{minipage}[b]{0.155\textwidth}
  \begin{overpic}[width=\linewidth]{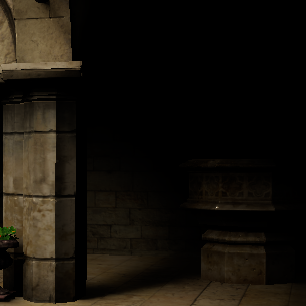}
  \put (53,2.5) {\textcolor{white}{\footnotesize \reflectbox{F}LIP: 0.05}}
  \put (21, 89) {\textcolor{white}{\footnotesize PG (Ours) +SVGF}}
 \end{overpic}
 \end{minipage}
 \\
 \begin{minipage}[b]{0.155\textwidth}
  \begin{overpic}[width=\linewidth]{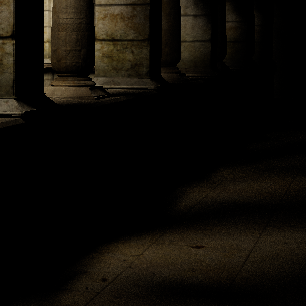}
 \put (4, 2.5) {\textcolor{white}{\footnotesize Reference}}
  \end{overpic}
  \end{minipage}
  \begin{minipage}[b]{0.155\textwidth}
 \begin{overpic}[width=\linewidth]{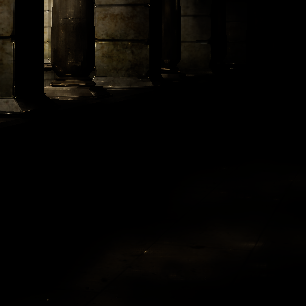}
 \put (53,2.5) {\textcolor{white}{\footnotesize \reflectbox{F}LIP: 0.10}}
  \put (51.5, 89) {\textcolor{white}{\footnotesize PT +SVGF}}
 \end{overpic}
  \end{minipage}
  \begin{minipage}[b]{0.155\textwidth}
  \begin{overpic}[width=\linewidth]{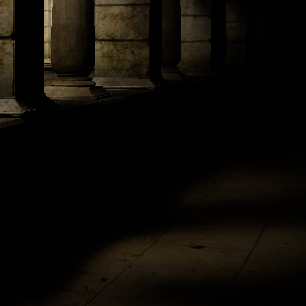}
  \put (53,2.5) {\textcolor{white}{\footnotesize \reflectbox{F}LIP: 0.03}}
  \put (21, 89) {\textcolor{white}{\footnotesize PG (Ours) +SVGF}}
 \end{overpic}
 \end{minipage}
   \\[-0.4cm]
 \begin{center}
    \scalebox{0.55}{\input{histogram.pgf}}
\end{center}
 \caption{\label{fig:firstExample}
           As can be seen in this comparison for the Sponza scene, applying real-time SVGF can achieve robust results by using our path-guiding technique. Relying only on the denoiser with default path-tracing signal leads to temporal-flickering, fireflies, over-darkening the final renders, and loss of high-frequency indirect lighting or shadows. Also, we attach the plot for visualizing the level of flickering, which is captured by rendering several frames of the static scene.}
 \end{figure}
\subsection{Performance}
Because we rely on screen space technique the performance is fully dependent on the resolution of frame buffers. For improving it we preloaded VPL from global memory to local shared memory in the second training pass, which allows us to sample \(N\) close VPL with better efficiency. The value of \(N\) influences the final performance and it introduces the trade-off between the speed of convergence of training and load on computing resources, that's why we propose to estimate \(N = (1.0-k/kMax)*15+5\), where \(kMax\) - maximum number of training epochs. The final results of our performance measurements are about 1.35-1.5 ms per frame on RTX 2070 at 1080p resolution as additional overhead compares to the default path-tracing technique. But at the same time, path-guiding promotes the generation of more coherent rays, that's why it leads to overall performance gain in about 5\% compare to default path-tracing on average in our experiments.
\section{Current Pitfalls}
One of the main problems of this algorithm - is slow training of the mixture model parameters in cases of incorrect reprojection between pixels which has very different incident lighting distribution. We discovered that it's not sufficient to cut pixels for re-projection only by normal and depth differences, so the optimal strategy - is an open question for future research. Also, using the screen space technique leads to unstable results for new visible surfaces, but we want to notice that convergence in such cases is very rapid.

Because of supporting guiding in the hemispherical domain, we rely only on BRDF distribution for transparent materials and also for materials with roughness near zero. 
\section{Conclusion}
In this paper, we have presented a novel real-time technique for path-guiding which decreases the overall noise levels of rendering with a limited samples budget for different types of scenes and light conditions, and slightly improves the overall performance. Also, we discovered that our algorithm has additional benefits of combining it with SVGF, which leads to decreasing the bias of the final render and improving its temporal stability. We believe that our work will inspire the research community to explore new ways of applying path-guiding techniques for real-time ray-tracing.
\section{Acknowledge}
We express our gratitude to Frolov Vladimir, Klepikov Dmitrii, Feklisov Egor, Alexander Veselov for helpling with copy-editing, reviewing and translating the paper.
\bibliographystyle{alpha-doi} 
\newcommand{\etalchar}[1]{$^{#1}$}


\end{document}